\documentclass[runningheads]{llncs}
\usepackage[T1]{fontenc}
\usepackage{graphicx}
\usepackage{multirow}
\usepackage{booktabs}
\newcommand{\bftab}{\fontseries{b}\selectfont}
\begin{document}
\title{Arabic ASR on the SADA Large-Scale Arabic Speech Corpus with Transformer-Based Models}
\titlerunning{Arabic ASR on SADA with Transformer models}
%
\author{Branislav Gerazov\inst{1}\orcidID{0000-0003-2498-6831} \and
Marcello Politi\inst{2}\orcidID{0009-0001-3195-5497} \and \\
Sébastien Bratières\inst{2}\orcidID{0000-0002-8117-0153} 
}
\authorrunning{B. Gerazov et al.}
%
\institute{
  FEEIT, Ss Cyril and Methodius University, Rugjer Boshkovikj 18, 1000 Skopje, Macedonia\\
  \email{gerazov@feit.ukim.edu.mk}
  \and
  Pi School, Via Indonesia 23, 00144 Rome, Italy\\
  \email{\{sebastien, marcello.politi\}@picampus-school.com}
}
\maketitle              
\begin{abstract}
We explore the performance of several state-of-the-art automatic speech recognition (ASR) models on a large-scale Arabic speech dataset, the SADA (Saudi Audio Dataset for Arabic), which contains 668 hours of high-quality audio from Saudi television shows.
The dataset includes multiple dialects and environments, specifically a noisy subset that makes it particularly challenging for ASR.
We evaluate the performance of the models on the SADA test set, and we explore the impact of fine-tuning, language models, as well as noise and denoising on their performance.
We find that the best performing model is the MMS 1B model finetuned on SADA with a 4-gram language model that achieves a WER of 40.9\% and a CER of 17.6\% on the SADA test clean set.

\end{abstract}

\section{Introduction}

Automatic Speech Recognition (ASR) is a technology that converts spoken language into text.
ASR systems have been around for decades, but they have seen a significant improvement in the last few years with the advent of deep learning and the creation of large-scale datasets.
The most notable improvements have been achieved with the introduction of end-to-end models, such as Wav2Vec2 \cite{baevski2020wav2vec} and Whisper \cite{radford2023robust}, which are based on Transformer architectures \cite{vaswani2017attention}.

In this work, we focus on Arabic ASR, which is a challenging task due to the complexity of the Arabic language, especially in terms of its dialectal variety.
Arabic is spoken by more than 400 million people worldwide, and it has many dialects that differ significantly from each other, as well as the Modern Standard Arabic (MSA) language norm.
It is also a language that has not been the mainstream focus of ASR research, with limited resources available for training and evaluating ASR systems.

In recent years, there has been growing interest in Arabic ASR, with several datasets and models being developed.
Dhouib et al. \cite{dhouib2022arabic} provide a systematic literature review of Arabic ASR, covering the period from 2011 to 2021 and focusing on the toolkits, datasets, and techniques used in Arabic ASR. They found that KALDI \cite{povey2011kaldi} and HTK \cite{young2002htk} were the most popular toolkits, and that 89.5\% of the studies focused on Modern Standard Arabic (MSA), while 26.3\% focused on various Arabic dialects.
Abdelhamid et al. \cite{abdelhamid2020end} in addition provide a review of the development of end-to-end Arabic ASR models.
Recently, Besdouri et al. \cite{besdouri2024arabic} provided a comprehensive overview of Arabic ASR, focusing on the challenges posed by the language's diverse forms and dialectal variations.
To address the challenges of dialectal code-switching in Arabic ASR, Chowdhury et al. \cite{chowdhury2021towards} propose a multilingual end-to-end ASR system based on the conformer architecture \cite{gulati2020conformer} that outperforms state-of-the-art monolingual dialectal Arabic and code-switching Arabic ASR systems.
Finally, the Open Universal Arabic ASR Leaderboard has been introduced to benchmark open-source general Arabic ASR models across various multi-dialect datasets \cite{wang2024open}.

We explore the performance of several state-of-the-art ASR models on a large-scale Arabic speech dataset, the SADA (Saudi Audio Dataset for Arabic) \cite{alharbi2024sada}, which contains 668 hours of high-quality audio from Saudi television shows.
The dataset includes multiple dialects and environments, specifically a noisy subset that makes it particularly challenging for ASR.
We evaluate the performance of the models on the SADA test set, and we explore the impact of fine-tuning, language models, as well as noise and denoising on the performance of these models.

\section{Dataset}

There are several datasets available for Arabic ASR. 
The Arabic Speech Corpus -- a single speaker dataset created for TTS in the Syrian Damask dialect, with some 2k samples \cite{halabi2016arabic}. 
The Egyptian-ASR-MGB-3 dataset of 16 h manually transcribed multi-genre data of Egyptian collected from different YouTube channels. 
The Tarteel Recitation Dataset of Quranic recitation with 67.4 h from 1,200 speakers \cite{khan2021tarteel}. 
Finally, Mozilla Common Voice \cite{ardila2019common} that as of v22 contains 157 h of Arabic (92 h validated) with 1,632 speakers.\footnote{
  \url{https://commonvoice.mozilla.org/en/datasets}
}


In our work, we focus on the SADA (Saudi Audio Dataset for Arabic) dataset -- a large-scale Arabic speech dataset with 668 hours of high-quality audio \cite{alharbi2024sada}. 
The audio is sourced from 57 Saudi Broadcasting Authority television shows, suitable for speech recognition training. 
It includes both read and spontaneous speech in multiple dialects  -- primarily the three major Saudi dialects (Najdi, Hijazi and Khaleeji), but also including Yemeni, Egyptian, and Levantine. 
The dataset was transcribed and prepared by the National Center for Artificial Intelligence in Saudi Arabia.

Our initial data exploration shows SADA to be a challenging dataset. 
There are quite long samples (>30 s), as well as issues with speaker overlap, and transcription errors.
Moreover, less than a third of the data is considered clean, with another third classified as noisy and the final one containing music, as shown from the data spread in Table~\ref{tab:sada}.
We see this as a good opportunity to test the performance of state-of-the-art ASR models on a challenging large-scale Arabic dataset, and to explore the impact of fine-tuning and denoising on the performance of these models.

\begin{table}
  \centering
  \caption{SADA data spread across the environments.}\label{tab:sada}
  {\setlength{\tabcolsep}{8pt}
  \begin{tabular}{lllll}
    \toprule
    Environment & Train (h) & Valid (h) & Test (h) & Total (h)\\
    \midrule
    Clean & 116.19 & 2.77 & 3.73 & 122.69\\
    Music & 159.29 & 4.13 & 3.17 & 166.59\\
    Noisy & 141.73 & 2.26 & 3.84 & 147.83\\
    Car & 0.42 & 0 & 0 & 0.42\\
    \midrule
    Total & 417.63 & 9.16 & 10.74 & 437.53\\
    \bottomrule
  \end{tabular}
  }
\end{table}

To explore the impact of noise on the performance of the ASR systems, we analyze more closely the contents of the noisy part of the SADA dataset.
It appears that most of the data is not that noisy, i.e. sometimes there is some laughter or English words present.
Otherwise, there are several type of noise:
\begin{itemize}
  \item audience noise (including chatter, laughter and applause),
  \item traffic noise,
  \item white/brown noise (sometimes low frequency),
  \item nature sounds, e.g. birds and dogs,
  \item harmonic noise, i.e. like slow music,
  \item audio effects, e.g. doors, thumping, objects hitting,
  \item microphone noise, i.e. microphone handling noise.
\end{itemize}

\section{Methodology}

We consider several state-of-the-art ASR models to build our ASR system.
We evaluate their performance on SADA w.r.t. WER (Word Error Rate) and CER (Character Error Rate).
We then explore improving them via fine-tuning, adding language models and denoising.

\subsection{ASR Models}
Our initial pick for is the XLSR-53 model \cite{conneau2020unsupervised} finetuned for Arabic by Mohamed El-Geish as a baseline model for our experiments, which we will refer to as simply \emph{XLSR}.\footnote{
  \url{https://huggingface.co/elgeish/wav2vec2-large-xlsr-53-arabic}
}
The base XLSR-53 model is a multilingual 300 M parameter Wav2Vec2 \cite{baevski2020wav2vec} model pretrained on 50 kh from 53 languages, no Arabic. 
The elgeish-xlsr model was finetuned on the Arabic Speech Corpus and Common Voice v6.1, which contains 78 h of Arabic (50 h validated) of 672 speakers.
The model works with 16 kHz sampled audio and includes no language model (LM).

Next, we consider XLS-R \cite{babu2021xls} -- a Wav2Vec2 model pretrained on 436 kh in 128 languages, including 95 h Arabic.
The XLS-R model comes in three model sizes of 300 M, 1 B and 2 B parameters.
We choose the 300 M parameter model for our experiments, which is the same size as elgeish-xlsr, and henceforth refer to it as.

Finally, we consider Whisper and MMS. 
Whisper \cite{radford2023robust} is a family of large transformer-based models pretrained on 680 kh of audio in 97 languages, including 739 h of Arabic. 
The models come in five sizes, from 39 M (tiny) to 1.55 B (large) parameters. 
The large model has been updated to v2 that features 2.5$\times$ more training epochs and added regularization, and v3 that is trained on a larger dataset of 5$+$ Mh of audio.
We did not use the large model as we chose to limit the analysis to models that can run on a single consumer grade GPU with 8 GB of VRAM.
Whisper's built-in neural sequence decoder acts in part like a language model.

MMS (Massively Multilingual Speech) \cite{pratap2023mms} is a 300 M and 1 B parameter Wav2Vec2 model pretrained on 1,406 languages with 491 kh of speech, and then finetuned for ASR in 1,107 languages using 44.7 kh of labeled speech data, and additional adapter layers (2 M parameters) that are finetuned for each language.
MMS $+$ LM outperforms Whisper medium and large v2, halving the WER and CER whilst being trained on more than 10$\times$ less labeled data and supporting 10$\times$ more languages. 

\subsection{Fine-tuning}

The largest improvements in ASR performance are achieved by fine-tuning the model on the target dataset.
We explore finetuning of the XLSR, XLS-R and MMS models on SADA.
In order to maintain a fair comparison, we finetune all models for 100k steps, which is the number of steps used by Elgeish to finetune the XLSR model on the Common Voice v6.1 Arabic dataset.\footnote{
  Note that Elgeish first finetuned the XLSR model on the Arabic Speech Corpus, but the number of steps is not disclosed.
}

For finetuning the XLSR and MMS models we experiment with unfreezing the encoder weights.
We also explore finetuning the original XLSR model and not the already finetuned El-Geish XLSR model.


\subsection{Language models}

Language models can improve the performance of ASR systems, especially in the absence of finetuning.
For example, adding a 4-gram language model to an English Wav2Vec2 ASR model, when the model has only been finetuned with 10 min of speech, decreases the WER on the Librispeech test clean set from 40.2\% to 6.6\% and adding a larger Trasnformer-based language model decreases it further to 4.8\% \cite{baevski2020wav2vec}.
In the case of finetuning, the language model leads only to a marginal improvement, i.e. from 2.2\% to 2.0\% WER and then to 1.8\% WER with a larger Transformer-based language model.

In our experiments we choose a 4-gram language model in order to avoid increasing the overall ASR system complexity, while reaping the benefits of a language model.
We choose KenLM -- an n-gram Language Model with Kneser-Ney smoothing, fast and low-memory querying \cite{heafield-2011-kenlm}, as well as fast and scalable estimation based on its streaming algorithms \cite{heafield-etal-2013-scalable},

\subsection{Noise and Denoising}

Since one third of the SADA data is noisy, we explore the impact of noise on the performance of the ASR systems in more detail.
We try finetuning the XLSR model on the noisy SADA data, and then evaluate its performance.

We also explore denoising and its potential to improve the performance of the ASR systems.
We focus on denoising using spectral gating, as implemented in the Noisereduce algorithm \cite{sainburg2020finding} that first computes a spectrogram of a signal and estimates a noise threshold (or gate) for each frequency band of that signal/noise. 
The threshold is then used to compute a mask, which gates noise below the frequency-varying threshold.

In its non-stationary version, the estimated noise threshold is continuously updated over time.
This relies on the fact that most types of noise occur at timescales larger than the timescale of the speech signal.
In practice, the spectrogram of the signal is time-smoothed and it is used to calculate a noise mask that is smoothed with a filter over frequency and time.

We use Noisereduce to denoise the SADA noisy data, and evaluate the XLSR model on the denoised data.
We also try finetuning the XLSR model on the denoised SADA data, and evaluate its performance on the denoised data.

Finally, we evaluate if denoising helps performance w.r.t. to the noise level.
Our hypothesis is that denoising helps more on noisier data, and hurts performance on less noisy data.
SADA is a good dataset for this analysis as its noisy set contains samples that span the range from being very noisy and not noisy at all.
We use two proxies for the noise in the data:
\begin{itemize}
  \item the CER, which we assume is inversely correlated with the noise level, and 
  \item the average relative noise spectrogram energy extracted with Noisereduce.
\end{itemize}
For both of these proxies we calculate the absolute improvement of CER due to denoising as the difference between the CER on the noisy and the CER on the denoised data.

\begin{table}[t]
  \centering
  \caption{
    ASR model performance on SADA clean test set.
  }\label{tab:results}
  {\setlength{\tabcolsep}{8pt}
  \begin{tabular}{llrrr}
    \toprule
    \multicolumn{2}{c}{Model} & Size & WER & CER \\
    \midrule
    \multirow{4}{*}{\shortstack[l]{Base \\ models}} & xlsr\_elgeish & 300 M & 93.75 & 53.91\\
    & Whisper small & 300 M & 254.86 & 215.27\\
    & Whisper medium & 770 M & 116.69 & 154.70\\
    & MMS & 1 B & \bftab 84.40 & \bftab 44.20\\
    \midrule
    \multirow{5}{*}{\shortstack[l]{$+$ Finetuning}} & xlsr\_elgeish\_sada & 300 M & 54.34 & 23.86\\
    & xlsr\_elgeish\_sada\_unfreeze & 300 M & 51.64 & 22.13\\
    & xlsr\_sada & 300 M & 91.47 & 51.38\\
    & xls-r\_sada & 300 M & 90.65 & 49.22\\
    & MMS\_unfreeze & 1 B & \bftab 51.49 & \bftab 19.90\\
    \midrule
    \multirow{2}{*}{\shortstack[l]{ $+$ Language \\ models}} & xlsr\_elgeish\_sada\_unfreeze\_lm & 300 M & 42.62 & 18.31\\
    & MMS\_unfreeze\_lm & 1 B & \bftab 40.86 & \bftab 17.60\\
    \bottomrule
\end{tabular}
  }
\end{table}

\section{Results}

The results from using the chosen ASR models on the SADA test clean dataset are reported in Table~\ref{tab:results}.
The table also shows the performance of the models finetuned on the SADA training set, as well as the performance of the models finetuned on the SADA training set with an unfreezed encoder.

\subsection{Base Models}
The results obtained when using the ASR models directly without finetuning confirm the intuition we gathered from the data exploration stage that SADA is a challenging dataset.
The El-Geish XLSR model gave a WER of almost 100\% and a CER of around 54\%, meaning that every word and every second character is wrong!

The two Whisper models -- small and medium, perform the worst with a WER of 254.9\% and 116.7\%, and a CER of 215.3\% and 154.7\%, respectively.
On closer inspection, we found that for some test samples the WER is up to 15000\%! 
We identified the root cause of these results to be the proclivity of the Whisper model to hallucinate and generate unbound text.
The medium size model tends to do less hallucination, explaining the better performance.
An example of this behavior can be seen in to following sample:

\begin{figure}[h!]
  \vspace{-2em}
  \centering
  \includegraphics[width=\textwidth]{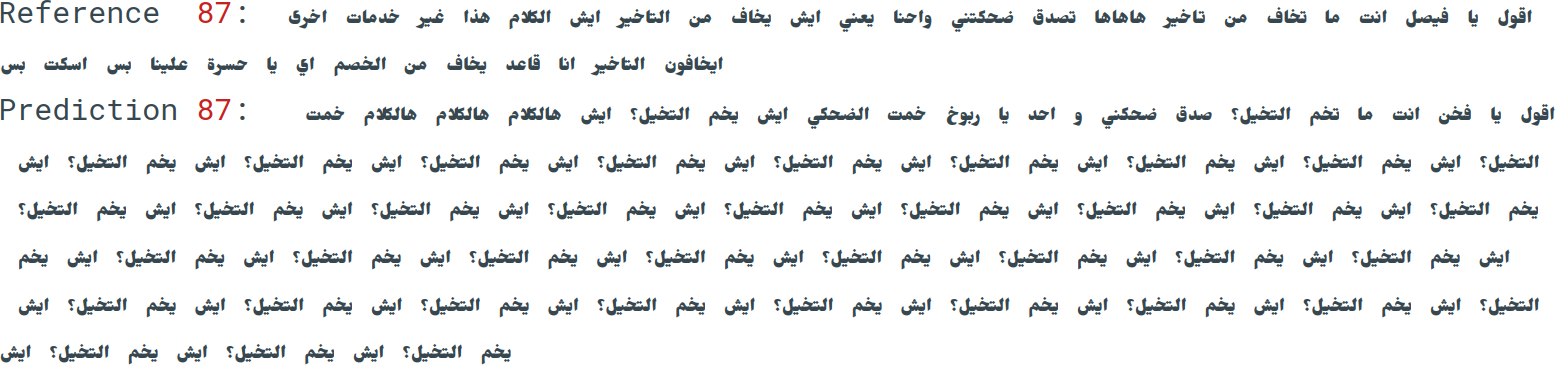}
  \label{fig:whisper_hallucination}
  \vspace{-3em}
\end{figure}

The MMS 1B model is the largest model we tested. 
It achieves the best performance with a WER of 84.4\% and a CER of 44.2\%, which is still quite high.

\begin{figure}[t]
  \centering
  \includegraphics[width=0.49\textwidth]{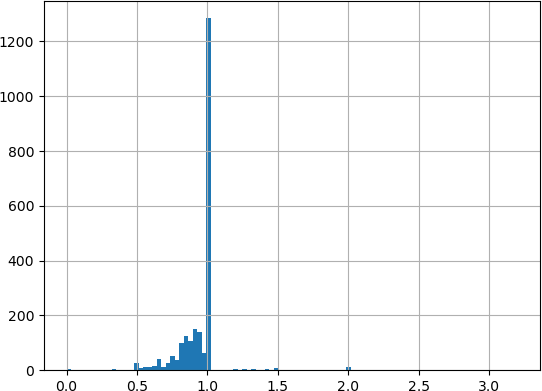}
  \hfill
  \includegraphics[width=0.49\textwidth]{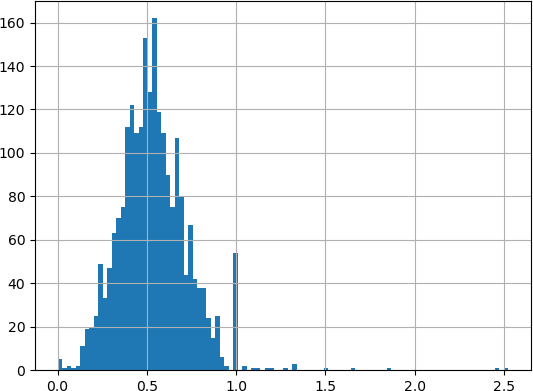}
  \vspace{-1em}
  \caption{WER (left) and CER histograms (right) with count on the \emph{y}-axis for the El-Geish XLSR model on the SADA test clean dataset.}
  \label{fig:wer_cer_hist}
\end{figure}

To get a clearer picture of the error spread across the test clean dataset, we plot a histogram of the WER and CER for the XLSR model in Fig.~\ref{fig:wer_cer_hist}.
We can clearly see that the mode for the WER distribution is 100\% and it dominates the distribution, with only a small tail going down to 50\%. 
The mode for the CER distribution is around 50\%, but the distribution is more spread out and looking like a Gaussian.

\subsection{Finetuning}

We finetune the ASR models on a subset of the SADA clean training set, which contains samples with lengths between 2~s and 10~s.
We set the lower bound based on the distribution of the CER for short samples, which falls mostly below 100\% CER for samples longer than 2~s. 
The upper bound is set to 10~s, to allow us to finetune the model on a 8 GB GPU.
The finetuning subset contains 35k of the 74k samples in the whole clean train set.

\subsubsection{XLSR}

\begin{figure}[t]
  \centering
  \includegraphics[width=0.49\textwidth]{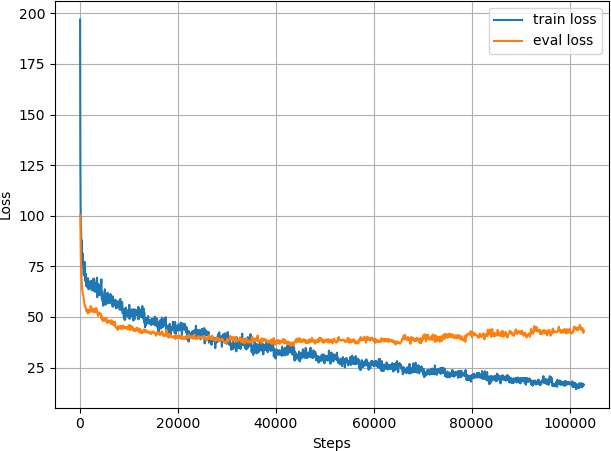}
  \hfill
  \includegraphics[width=0.49\textwidth]{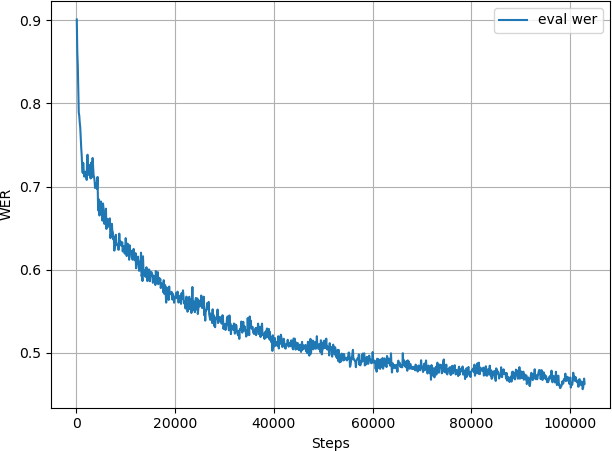}
  \vspace{-1em}
  \caption{Finetuning progress of the XLSR model on the SADA clean training set.}
  \label{fig:finetuning_progress}
  \vspace{-1em}
\end{figure}

The progress of finetuning the XLSR model on the SADA clean training set is shown in Fig.~\ref{fig:finetuning_progress}.
We can see that even though the validation (eval) loss starts increasing after 43k steps, the validation WER keeps decreasing.
We can also see that there is still room for improvement after 100k steps of finetuning.

\begin{figure}[t]
  \centering
  \includegraphics[width=0.49\textwidth]{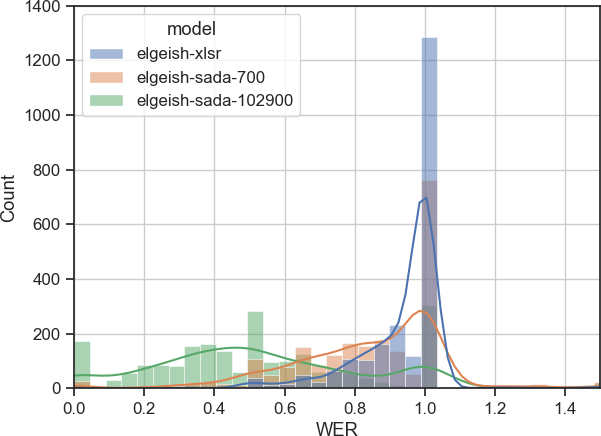}
  \hfill
  \includegraphics[width=0.49\textwidth]{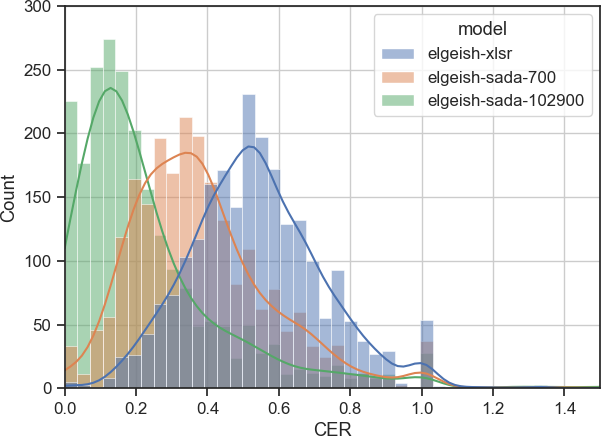}
  \vspace{-1em}
  \caption{WER (left) and CER histograms (right) with count on the \emph{y}-axis for the El-Geish XLSR model on the SADA test clean dataset after finetuning with 700 and 100k steps.}
  \label{fig:wer_cer_hist_finetuned}
  \vspace{-1em}
\end{figure}

The results for the finetuned XLSR model show a substantial improvement in performance, with WER going from 93.75\% down to 54.3\% and CER from 53.91\% down to 23.9\%, which is an improvement of  over 40\%.
We can see the changes of the distributions of the WER and CER with finetuning in Fig.~\ref{fig:wer_cer_hist_finetuned}. 
The plots also show the performance at 700 steps of finetuning, which we can see for the CER is half way to the final distribution for 100k steps of finetuning, exemplifying the law of diminishing returns.

\begin{figure}[t]
  \centering
  \includegraphics[width=0.49\textwidth]{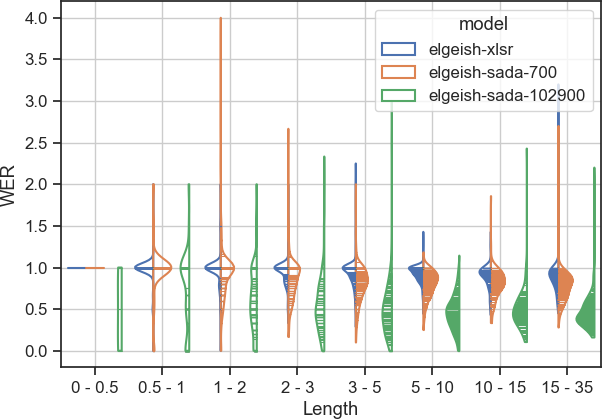}
  \hfill
  \includegraphics[width=0.49\textwidth]{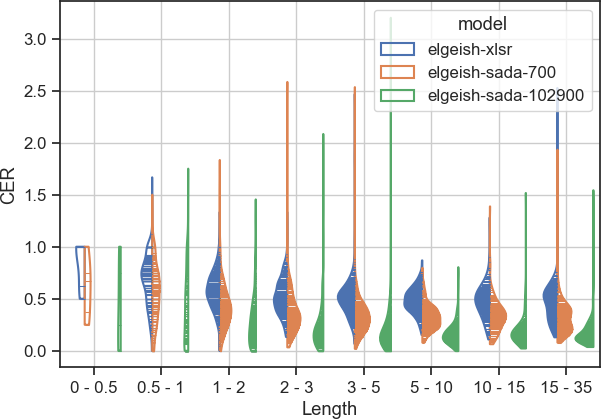}

  \vspace{1em}
  \includegraphics[width=0.49\textwidth]{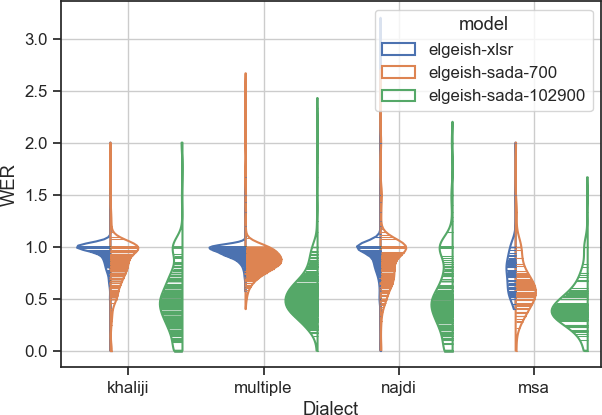}
  \hfill
  \includegraphics[width=0.49\textwidth]{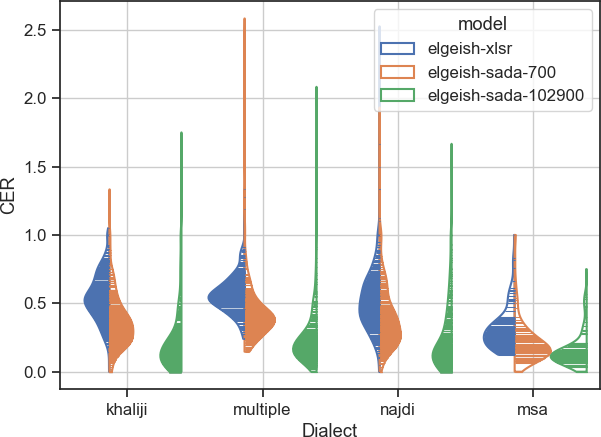}
  \vspace{-1em}
  \caption{WER (left column) and CER (right column) for the El-Geish XLSR model on the SADA test clean dataset after finetuning with 700 and 100k steps, spread across sample length (top row) and dialect (bottom row).}
  \label{fig:wer_cer_finetuning}
  \vspace{-1em}
\end{figure}

In Fig.~\ref{fig:wer_cer_finetuning} we break down the WER and CER performance of the finetuned XLSR model w.r.t. the sample length and the dialect.
We can see that the model performs better on longer samples and struggles with samples shorter than 3 s.
We can also see that finetuning improves this performance, less so for the short samples.

With respect to the dialect, we can see that the original model performs markedly better on MSA than on the Khaliji and Najdi dialects.
This can be expected as El-Geish finetuned XLSR on Damascian via Arabic Speech Database and on MSA via Common Voice.
Finetuning significantly improves performance for unseen dialects, with MSA still performing best over all.

This gives a marginal improvement to performance giving a WER of 51.6\% and a CER of 22.1\%, which is an further improvement of 2.7\% WER and 1.7\% CER over the finetuned XLSR model.

Finally, finetuning the original XLSR model, instead of the already finetuned El-Geish XLSR model, gives a WER of 91.5\% and a CER of 51.4\%, which is a slight improvement of 2.3\% WER and 1.5\% CER over the El-Geish XLSR model, but still worse than the finetuned El-Geish XLSR model. 
This is probably due to the El-Geish XLSR model being already finetuned for at least 100k steps, meaning that our finetuning brings it to 200k steps, double the number of steps we used for finetuning the original XLSR model.

\subsubsection{XLS-R}

\begin{figure}[t]
  \centering
  \includegraphics[width=0.49\textwidth]{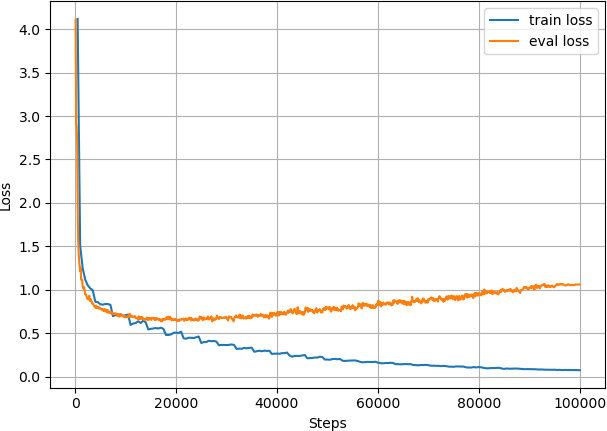}
  \hfill
  \includegraphics[width=0.49\textwidth]{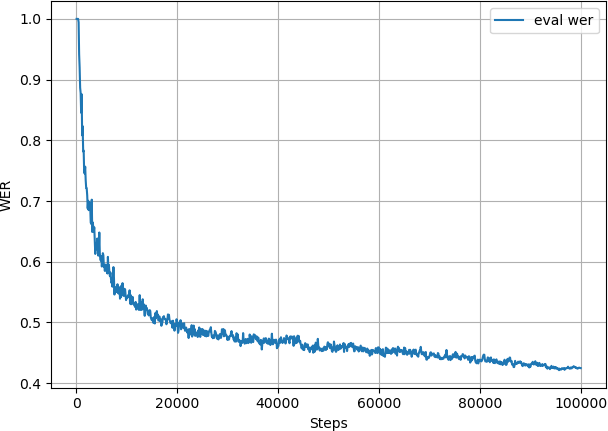}
  \vspace{-1em}
  \caption{Finetuning progress of the XLS-R model on the SADA clean training set.}
  \label{fig:finetuning_xlsr_loss}
  \vspace{-1em}
\end{figure}

The loss curves from finetuning XLS-R on SADA are shown in Fig.~\ref{fig:finetuning_xlsr_loss}.
We can see that the validation loss starts increasing after 20k steps, but the validation WER keeps decreasing.

The finetuned XLS-R model shows worse results than the finetuned on SADA El-Geish XLSR model, with a WER of 90.7\% down and CER of 49.2\%.\footnote{
  We also tried continued pretraining of XLS-R on SADA before finetuning, but the contrastive and eval loss curves showed that the process is failing.
}


\begin{figure}[t]
  \centering
  \includegraphics[width=0.49\textwidth]{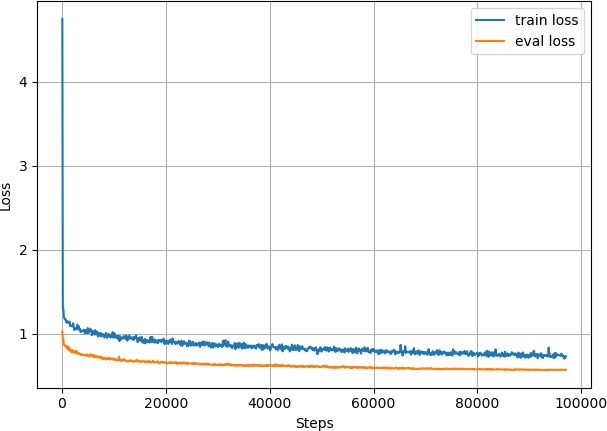}
  \hfill
  \includegraphics[width=0.49\textwidth]{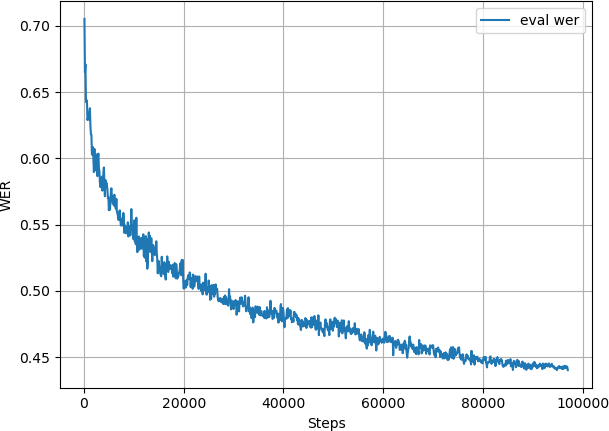}
  \vspace{-1em}
  \caption{Finetuning progress of the MMS model on the SADA clean training set.}
  \label{fig:finetuning_mms_loss}
  \vspace{-1em}
\end{figure}

\subsubsection{MMS} 
\begin{figure}[t]
  \centering
  \includegraphics[width=0.49\textwidth]{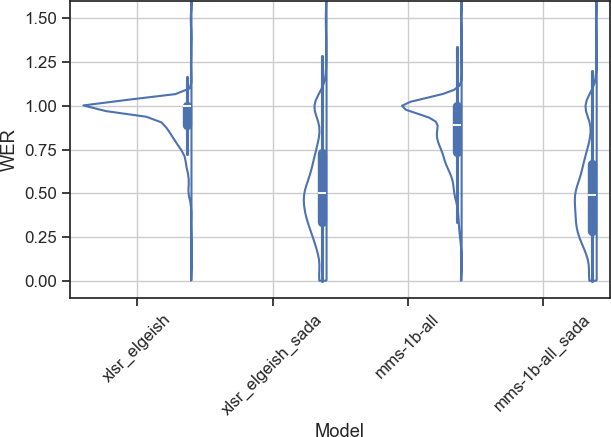}
  \hfill
  \includegraphics[width=0.49\textwidth]{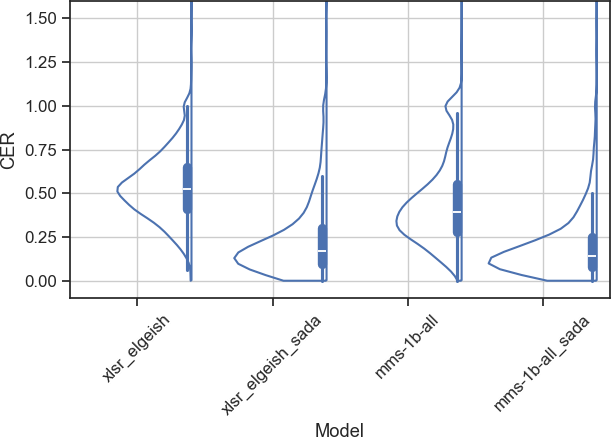}
  \vspace{-1em}
  \caption{WER (left) and CER (right) distributions for the El-Geish XLSR and MMS models and their versions finetuned on SADA.}
  \label{fig:wer_cer_hist_finetuned_mms}
\end{figure}

The finetuning progress of the MMS 1B model on SADA is shown in Fig.~\ref{fig:finetuning_mms_loss}.
We can see that both the validation loss and WER keep decreasing, even though they seem to near convergence.
The finetuned MMS 1B model gives a WER of 51.5\% and a CER of 19.9\%, which is the best performance we achieved on the SADA clean test set with finetuning for all the considered ASR models.
The distribution of the WER and CER for the best-performing finetuned models is shown in Fig.~\ref{fig:wer_cer_hist_finetuned_mms}.

\subsection{Language Models}

We train a 4-gram language model on the SADA training set using KenLM.
The obtained ARPA file is 2.2 million lines long (3 M for a 5-gram model) and a size of 100 MB. 
We integrate the language model into the XLSR and MMS models and obtain a absolute WER improvement of 9 and 11\% and a CER improvement of 4 and 2\%.
With this we achieve the best performing ASR system on SADA with a WER of 40.9\% and a CER of 17.6\% for MMS 1B finetuned on SADA with an LM.

\subsection{Noise and Denoising}

\begin{table}
  \centering
  \caption{ASR model performance on SADA in noise and denoising.}\label{tab:results_noise}
  {\setlength{\tabcolsep}{8pt}
  \begin{tabular}{llll}
    \toprule
    Model & Domain & WER & CER \\
    \midrule
    xlsr\_elgeish & clean & 93.75 & 53.91\\
    xlsr\_elgeish & noisy & 96.87 & 58.27\\
    xlsr\_elgeish & denoised & 97.29 & 59.14\\
    \midrule
    xlsr\_elgeish\_sada & clean & 54.34 & 23.86\\
    xlsr\_elgeish\_sada & noisy & 63.43 & 29.63\\
    xlsr\_elgeish\_sada & denoised & 66.11 & 31.58\\
    \midrule
    xlsr\_elgeish\_sada\_unfreeze & clean & 51.64 & 22.13\\
    xlsr\_elgeish\_sada\_unfreeze & noisy & 60.95 & 28.09\\
    xlsr\_elgeish\_sada\_unfreeze & denoised & 64.46 & 30.47\\
    xlsr\_elgeish\_sada\_noisy\_unfreeze & noisy & \bftab 58.95 & \bftab 27.10\\
    xlsr\_elgeish\_sada\_denoised\_unfreeze & denoised & 59.13 & 28.85\\
    \bottomrule
  \end{tabular}
  }
\end{table}

The results from the evaluation of the XLSR model on the noisy subset are given in Table~\ref{tab:results_noise}. 
The results from the evaluation on the clean dataset are added for reference.
We can see that the models, including the ones finetuned on the clean SADA train set, perform worse on the noisy data.
The best model obtained with finetuning XLSR with unfreezing the encoder weights \emph{xlsr\_elgeish\_sada\_unfreeze} gives a WER of 60.9\% and a CER of 28.1\% on the noisy data, which is 9.3\% worse WER and 6\% worse CER than on the clean data.
Finetuning the model on the noisy SADA train set does improve performance and gives a WER of 58.9\% and a CER of 27.1\%, which is the best performance on the noisy data.
We note that this is still much worse than the performance on the clean data.

\begin{figure}[t]
  \centering
  \includegraphics[width=0.49\textwidth]{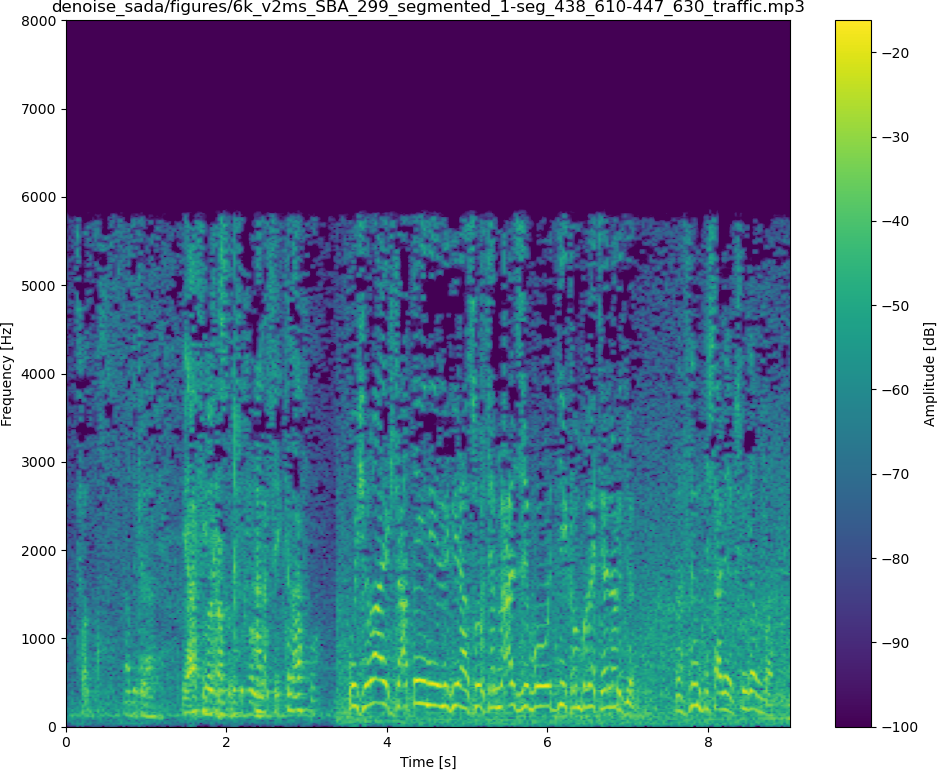}
  \hfill
  \includegraphics[width=0.49\textwidth]{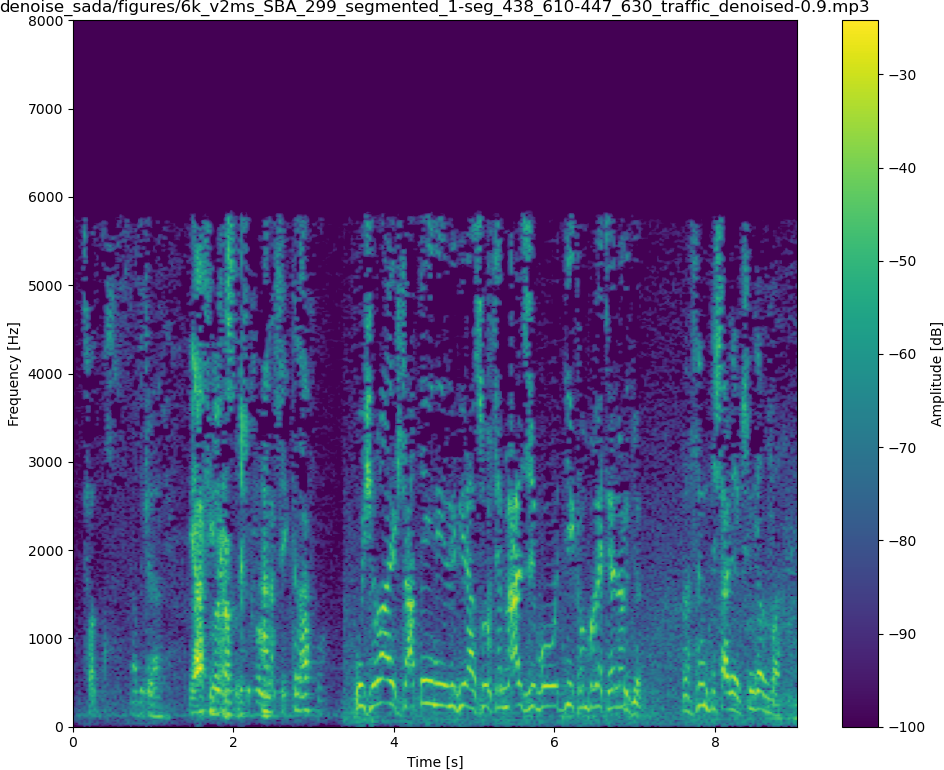}

  \includegraphics[width=0.49\textwidth]{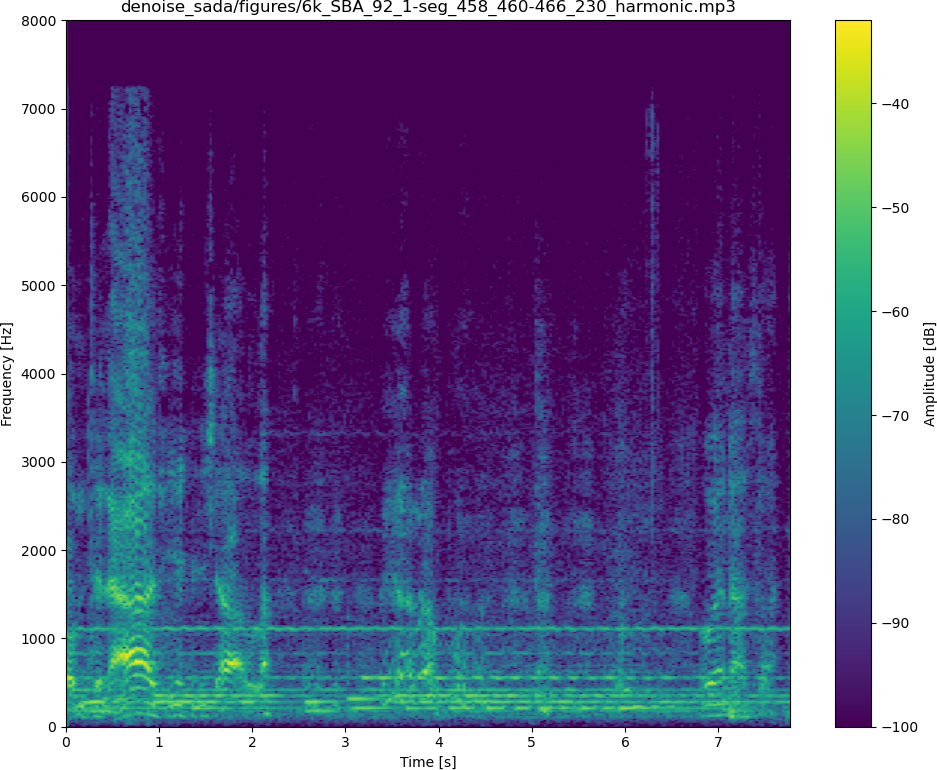}
  \hfill
  \includegraphics[width=0.49\textwidth]{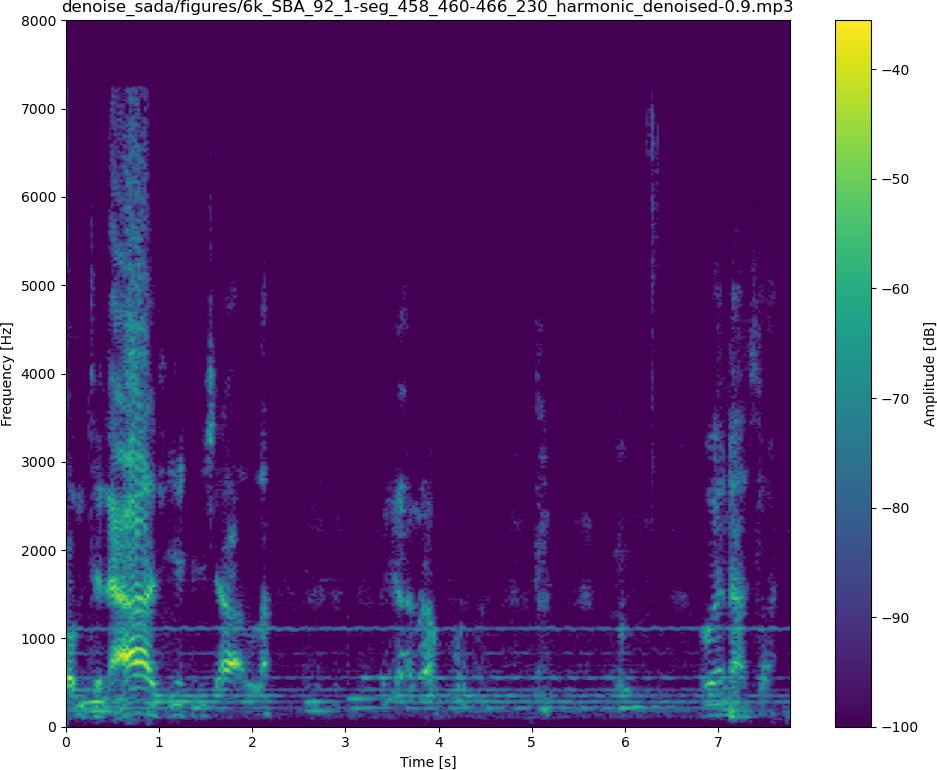}
  \caption{Example denoising results for traffic noise (top row) and harmonic noise (bottom row).}
  \label{fig:denoise_example}
\end{figure}

We next apply the Noisereduce denoising algorithm to the noisy SADA data.
Example results of the denoising process can be seen for traffic and harmonic noise in Fig.~\ref{fig:denoise_example}.
We set the noise reduction coefficient to 0.9, i.e.90\%, to avoid severely damaging the speech signal integrity.

From Table~\ref{tab:results_noise} we can see that denoising the noisy data does not help the XLSR model performance.
In fact, it is always detrimental to the performance of the system.
Even the best performing model, \emph{xlsr\_elgeish\_sada\_unfreeze}, obtains 3.5\% worse WER and 2.4\% worse CER than on the noisy data.
Finetuning the model to the denoised SADA dataset does improve performance, but it is still worse than finetuning and evaluating the model on the noisy data sans denoising.

\begin{figure}[t]
  \centering
  \includegraphics[width=0.49\textwidth]{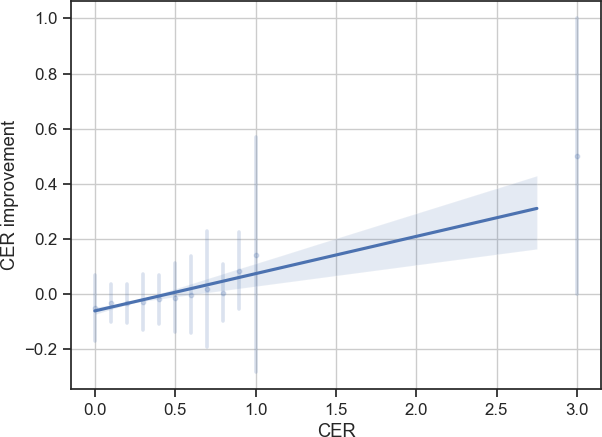}
  \hfill
  \includegraphics[width=0.49\textwidth]{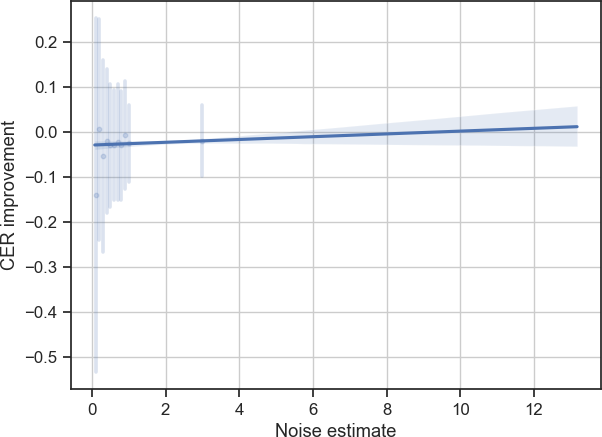}
  \caption{Absolute improvement of CER due to denoising w.r.t. the CER (left) and the Noisereduce average relative noise spectrogram energy (right).}
  \label{fig:denoise_cer_noise}
\end{figure}

Finally, we plot the absolute improvement of CER due to denoising w.r.t. the CER and the Noisereduce average relative noise spectrogram energy in Fig.~\ref{fig:denoise_cer_noise}.
We bin the data for clearer visualization and compute the average CER and noise estimate for each bin.
We fit a regression line to the data and compute the Pearson correlation coefficient, which equals 0.28 for the CER and 0.03 for the noise estimate.

We can see that denoising helps for higher CER, and hurts performance for lower CER. 
On average it leads to CER improvement if the CER is above 50\%.
The same is true for the noise estimate, i.e. denoising helps for higher noise.

\section{Conclusion}

We explored the performance of state-of-the-art ASR models on the SADA Arabic dataset.
We found that the 1B parameter MMS model finetuned on SADA with a 4-gram language model gives the best performance with a WER of 40.9\% and a CER of 17.6\%.
The 300M parameter XLSR model finetuned on SADA with an unfreezed encoder and a 4-gram language model gives a competitive WER of 42.6\% and a CER of 18.3\%, offering a good trade-off between performance and model size.
We also explored the impact of noise on the performance on the smaller XLSR model. 
We found that it performs worse on the noisy data, and that finetuning it on the noisy data is the best path towards improving performance.
Denoising the noisy data mostly does not help performance. 
Although, it does have the potential to improve performance on data with pronounced noise levels.

\begin{credits}
\subsubsection{\ackname} This study was funded by Pi School, Rome, Italy. The authors would like to thank the Pi School for providing the necessary resources and support for this research.
\end{credits}

\newpage
\bibliographystyle{splncs04}
\bibliography{refs}

\begin{thebibliography}{10}
\providecommand{\url}[1]{\texttt{#1}}
\providecommand{\urlprefix}{URL }
\providecommand{\doi}[1]{https://doi.org/#1}

\bibitem{abdelhamid2020end}
Abdelhamid, A.A., Alsayadi, H.A., Hegazy, I., Fayed, Z.T.: End-to-end arabic
  speech recognition: A review. In: Proceedings of the 19th Conference of
  Language Engineering. pp. 26--30 (2020)

\bibitem{alharbi2024sada}
Alharbi, S., Alowisheq, A., T{\"u}ske, Z., Darwish, K., Alrajeh, A., Alrowithi,
  A., Tamran, A.B., Ibrahim, A., Aloraini, R., Alnajim, R., et~al.: Sada: Saudi
  audio dataset for arabic. In: ICASSP 2024-2024 IEEE International Conference
  on Acoustics, Speech and Signal Processing (ICASSP). pp. 10286--10290. IEEE
  (2024)

\bibitem{ardila2019common}
Ardila, R., Branson, M., Davis, K., Henretty, M., Kohler, M., Meyer, J.,
  Morais, R., Saunders, L., Tyers, F.M., Weber, G.: Common voice: A
  massively-multilingual speech corpus. LREC  (2020)

\bibitem{babu2021xls}
Babu, A., Wang, C., Tjandra, A., Lakhotia, K., Xu, Q., Goyal, N., Singh, K.,
  Von~Platen, P., Saraf, Y., Pino, J., et~al.: Xls-r: Self-supervised
  cross-lingual speech representation learning at scale. arXiv preprint
  arXiv:2111.09296  (2021)

\bibitem{baevski2020wav2vec}
Baevski, A., Zhou, Y., Mohamed, A., Auli, M.: wav2vec 2.0: A framework for
  self-supervised learning of speech representations. Advances in neural
  information processing systems  \textbf{33},  12449--12460 (2020)

\bibitem{besdouri2024arabic}
Besdouri, F.Z., Zribi, I., Belguith, L.H.: Arabic automatic speech recognition:
  challenges and progress. Speech Communication  \textbf{163},  103110 (2024)

\bibitem{chowdhury2021towards}
Chowdhury, S.A., Hussein, A., Abdelali, A., Ali, A.: {Towards one model to rule
  all: Multilingual strategy for dialectal code-switching Arabic ASR}. arXiv
  preprint arXiv:2105.14779  (2021)

\bibitem{conneau2020unsupervised}
Conneau, A., Baevski, A., Collobert, R., Mohamed, A., Auli, M.: Unsupervised
  cross-lingual representation learning for speech recognition. arXiv preprint
  arXiv:2006.13979  (2020)

\bibitem{dhouib2022arabic}
Dhouib, A., Othman, A., El~Ghoul, O., Khribi, M.K., Al~Sinani, A.: Arabic
  automatic speech recognition: A systematic literature review. Applied
  Sciences  \textbf{12}(17), ~8898 (2022)

\bibitem{gulati2020conformer}
Gulati, A., Qin, J., Chiu, C.C., Parmar, N., Zhang, Y., Yu, J., Han, W., Wang,
  S., Zhang, Z., Wu, Y., et~al.: Conformer: Convolution-augmented transformer
  for speech recognition. arXiv preprint arXiv:2005.08100  (2020)

\bibitem{halabi2016arabic}
Halabi, N., et~al.: Arabic speech corpus. Oxford Text Archive Core Collection
  (2016)

\bibitem{heafield-2011-kenlm}
Heafield, K.: {K}en{LM}: Faster and smaller language model queries. In:
  Proceedings of the Sixth Workshop on Statistical Machine Translation. pp.
  187--197. Association for Computational Linguistics, Edinburgh, Scotland (Jul
  2011)

\bibitem{heafield-etal-2013-scalable}
Heafield, K., Pouzyrevsky, I., Clark, J.H., Koehn, P.: Scalable modified
  {K}neser-{N}ey language model estimation. In: Proceedings of the 51st Annual
  Meeting of the Association for Computational Linguistics (Volume 2: Short
  Papers). pp. 690--696. Association for Computational Linguistics, Sofia,
  Bulgaria (Aug 2013)

\bibitem{khan2021tarteel}
Khan, H.I., Abid, A., Moussa, M.M., Abou-Allaban, A.: {The Tarteel dataset:
  crowd-sourced and labeled Quranic recitation}  (2021)

\bibitem{povey2011kaldi}
Povey, D., Ghoshal, A., Boulianne, G., Burget, L., Glembek, O., Goel, N.,
  Hannemann, M., Motlicek, P., Qian, Y., Schwarz, P., et~al.: The kaldi speech
  recognition toolkit. In: IEEE 2011 workshop on automatic speech recognition
  and understanding. vol.~1, pp.~5--1. Hawaii (2011)

\bibitem{pratap2023mms}
Pratap, V., Tjandra, A., Shi, B., Tomasello, P., Babu, A., Kundu, S., Elkahky,
  A., Ni, Z., Vyas, A., Fazel-Zarandi, M., Baevski, A., Adi, Y., Zhang, X.,
  Hsu, W.N., Conneau, A., Auli, M.: Scaling speech technology to 1,000+
  languages. arXiv  (2023)

\bibitem{radford2023robust}
Radford, A., Kim, J.W., Xu, T., Brockman, G., McLeavey, C., Sutskever, I.:
  Robust speech recognition via large-scale weak supervision. In: International
  conference on machine learning. pp. 28492--28518. PMLR (2023)

\bibitem{sainburg2020finding}
Sainburg, T., Thielk, M., Gentner, T.Q.: Finding, visualizing, and quantifying
  latent structure across diverse animal vocal repertoires. PLoS computational
  biology  \textbf{16}(10),  e1008228 (2020)

\bibitem{vaswani2017attention}
Vaswani, A., Shazeer, N., Parmar, N., Uszkoreit, J., Jones, L., Gomez, A.N.,
  Kaiser, {\L}., Polosukhin, I.: Attention is all you need. Advances in neural
  information processing systems  \textbf{30} (2017)

\bibitem{wang2024open}
Wang, Y., Alhmoud, A., Alqurishi, M.: Open universal arabic asr leaderboard.
  arXiv preprint arXiv:2412.13788  (2024)

\bibitem{young2002htk}
Young, S., Evermann, G., Gales, M., Hain, T., Kershaw, D., Liu, X., Moore, G.,
  Odell, J., Ollason, D., Povey, D., et~al.: The htk book. Cambridge university
  engineering department  \textbf{3}(175), ~12 (2002)

\end{thebibliography}
\end{document}